\documentclass[twocolumn,superscriptaddress,floatfix,showpacs]{revtex4-1}
\usepackage{graphicx}
\usepackage{rotating}
\usepackage{amsfonts,amsmath,color}
\usepackage{amssymb}
\usepackage{psfrag}

\newcommand{\bs}{{\bf {s}}}
\newcommand{\br}{{\bf {r}}}
\newcommand{\bv}{{\bf {v}}}
\newcommand{\Reyn}{\mbox{\rm Re}_n}         
\newcommand{\bk}{{\bf {k}}}

\newcommand{\bom}{{\mbox{\boldmath $\omega$}}}

\begin{document}

\title{Vortex-density fluctuations, energy spectra and vortical 
regions in superfluid turbulence}

\author{Andrew W. Baggaley}
\email{andrew.baggaley@gla.ac.uk}
\affiliation{
School of Mathematics and Statistics, University of Glasgow,
Glasgow, G12 8QW, UK
}
\affiliation{
Joint Quantum Centre Durham-Newcastle, 
School of Mathematics and Statistics, Newcastle University,
Newcastle upon Tyne, NE1 7RU, UK
}
\author{Jason Laurie}
\affiliation{
Laboratoire de Physique, \'{E}cole Normale Sup\'{e}rieure de Lyon, 
46 all\'{e}e d'Italie, 69007, Lyon, France
}
\author{Carlo F. Barenghi}
\affiliation{
Joint Quantum Centre Durham-Newcastle, 
School of Mathematics and Statistics, Newcastle University,
Newcastle upon Tyne, NE1 7RU, UK
}
\begin{abstract}
Measurements of the energy spectrum and of the vortex-density fluctuation
spectrum in superfluid turbulence seem to contradict each
other. Using a numerical model, we
show that at each instance of time the total vortex line density
can be decomposed into two parts: one formed by metastable
bundles of coherent vortices, and one in which the vortices are randomly
oriented. We show that the former is responsible for the observed
Kolmogorov energy spectrum, and the latter for the spectrum of the vortex line density fluctuations. 
\end{abstract}
\maketitle


Below a critical temperature, liquid helium becomes a two-fluid system
in which an inviscid  superfluid component 
coexists with a viscous normal fluid component.
The flow of the superfluid is 
irrotational: superfluid vorticity is confined to vortex lines 
of atomic thickness around which the circulation takes 
a fixed value $\kappa$ (the quantum of circulation). 
Superfluid turbulence \cite{Vinen-Niemela,Skrbek-Sreeni} is easily created
by stirring either helium isotope ($^4$He or $^3$He-B),
and consists of a tangle of reconnecting
vortex filaments which interact with each other and 
with the viscous normal fluid (which may be laminar or turbulent).
The most important observable quantity is the vortex line density 
$L$ (vortex length per unit volume), 
from which one infers the average
distance between vortex lines, $\ell \approx L^{-1/2}$. 
Our interest is in the properties of superfluid turbulence and
their similarities with ordinary turbulence.

Experiments \cite{Maurer,Salort} have revealed that, 
if the superfluid turbulence is driven by grids or propellers, 
the distribution of 
the turbulent kinetic energy over length scales larger than $\ell$
obeys the celebrated $k^{-5/3}$ Kolmogorov scaling
observed in ordinary (classical) turbulence. Here $k$ is
the magnitude of the three-dimensional wavenumber
(wavenumber and frequency are related by $k=f/{\bar v}$,
where ${\bar v}$ is the mean flow).
Numerical calculations performed using either the
vortex filament model \cite{Araki,Baggaley-fluctuations} 
or the Gross-Pitaevskii equation \cite{Nore,Kobayashi} confirm the
Kolmogorov scaling. 
It is thought that the effect arises from the
partial polarization of the
vortex lines  \cite{Vinen-Niemela,Skrbek-Sreeni,Lvov},
but such effect has never been clearly identified.
Another important experimental observation is that
in both $^4$He \cite{Roche2007} and $^3$He-B \cite{Bradley2008}, 
the frequency spectrum of the fluctuations of $L$ 
has a decreasing $f^{-5/3}$ scaling
typical of passive objects \cite{Roche-Barenghi,Baggaley-fluctuations} 
advected by a turbulent flow.
This latter result seems to contradict the interpretation 
of $L$ as a measure of superfluid vorticity, $\omega=\kappa L$ 
which is usually made in the literature
\cite{Stalp1999,Golov2008,Skrbek2010,Bradley2008,Vinen-Niemela,Skrbek-Sreeni}.

In fact, from dimensional analysis, the vorticity spectrum corresponding to the
Kolmogorov law should increase with $f$ (as $f^{1/3}$), not decrease.
Since the vortex line density is a positive quantity, a better
analogy is to the enstrophy spectrum: however 
in classical turbulence this spectrum is essentially flat \cite{Ishihara,Zhou},
in disagreement with the helium experiments~\cite{Roche2007,Bradley2008}.

The aim of this letter is to reconcile these two sets of observations 
(each separately backed by numerical simulations). 
We shall show that, at any instant, the vortex tangle 
can be decomposed into two
parts: vortex lines which are locally polarised in the same direction,
forming metastable coherent bundles, and vortex lines which are randomly
oriented in space. The former is responsible for the 
Kolmogorov energy spectrum, and the latter for frequency
spectrum of the vortex line density.


Following Schwarz \cite{Schwarz}, we model vortex filaments
as space curves $\bs(\xi,t)$ which move according to

\begin{equation}
\frac{d{\bf s}}{dt}=\bv_s+\alpha \bs' \times (\bv_n-\bv_s)
-\alpha' \bs' \times \left(\bs' \times \left(\bv_n-\bv_s\right)\right),
\label{eq:Schwarz}
\end{equation}

\noindent
where $t$ is time, $\alpha$ and $\alpha'$ are known
temperature dependent friction
coefficients \cite{Donnelly-Barenghi}, $\bs'=d\bs/d\xi$ is the unit
tangent vector at the point $\bs$, $\xi$ is arc length, and
$\bv_n$ is the normal fluid velocity at the point $\bf s$.
We set the temperature to $T=1.9~\rm K$, typical of many finite temperature studies
(corresponding to $\alpha=0.206$ and $\alpha'=0.0083$).
The self-induced velocity
of the vortex line at the point $\bs$ is given by
the Biot-Savart law \cite{Saffman}
\begin{equation}
\bv_s (\bs,\,t)=
-\frac{\kappa}{4 \pi} \oint_{\cal L} \frac{(\bs-\br) }
{\vert \bs - \br \vert^3}
\times {\bf d}\br,
\label{eq:BS}
\end{equation}

\noindent
where $\kappa=9.97 \times 10^{-4}~\rm cm^2/s$
(in $^4$He) and the line integral
extends over the entire vortex configuration ${\cal L}$. 
The calculation is performed in a periodic cube of size $D=0.1~\rm cm$.
The numerical techniques to discretize the vortex lines
into a number of points $\bs_j$ ($j=1, \cdots N$) held
at minimum separation $\Delta\xi/2$, compute the time
evolution, de-singularize the Biot-Savart integrals, evaluate $\bv_s$
using a tree-method (with critical opening angle $0.4$), and 
algorithmically perform vortex reconnections when vortex lines
come sufficiently close to each other, are all
described in our previous 
papers~\cite{Baggaley-fluctuations,Baggaley-reconnections}.
Here we take $\Delta \xi=2 \times 10^{-3}~$cm and  a
timestep of $5 \times 10^{-5}~$s. The number of discretization points
varies with time; in the statistically steady state we focus on,
$N \approx 4000$.

The turbulent normal fluid is modelled by
the following synthetic turbulent flow \cite{Osborne}:

\begin{equation}
\bv_n(\bs,\,t)=\sum_{m=1}^{M}({\bf A}_m \times \bk_m \cos{\phi_m}
+{\bf B}_m \times \bk_m \sin{\phi_m}),
\label{eq:KS}
\end{equation}

\noindent
where $\phi_m=\bk_m \cdot \bs + f_m t$, $\bk_m$ and 
$f_m=\sqrt{k^3_m E(k_m)}$ are  wave vectors and angular frequencies.
This flow  is solenoidal and time-dependent;
with a suitable random choice of ${\bf A}_m$, ${\bf B}_m$  and $\bk_m$
(adapted to the periodic box). The normal fluid's energy spectrum
has Kolmogorov form $E(k_m)\sim k_m^{-5/3}$
in the range from $k_1$ (corresponding to the integral scale) to $k_M$
(corresponding to the dissipation scale). 
This synthetic model of turbulence compares very well with Lagrangian 
statistics obtained in direct numerical simulations of the Navier-Stokes
equations and experiments. 
Here we take $M=188$, $k_1=0.1~\rm cm$ and $k_M=1.8 \times 10^{-3}~\rm cm$, 
which corresponds to the Reynolds number $\Reyn=(k_1/k_M)^{4/3}\approx 200$.
For computational simplicity our model ignores any back-reaction of the 
vortex lines onto the normal fluid.


\begin{figure}
\begin{center}
\psfrag{L}{$L$}
\psfrag{t}{$t$}
\includegraphics[width=\linewidth,keepaspectratio]{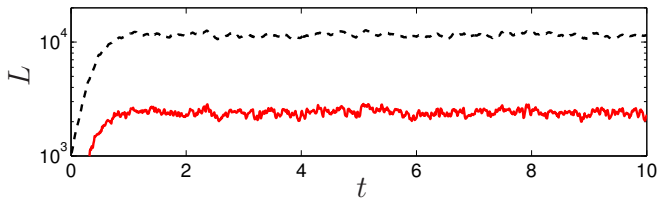}
\end{center}
\caption{(Color online) The time evolution of the 
total vortex line density (black dashed line) $L$ ($\rm cm^{-2}$) and the polarised component $L_{\parallel}$ ($\rm cm^{-2}$) (solid red line). 
}
\label{fig1}
\vspace{-4mm}
\end{figure}


We integrate the vortex lines in time according to Eq.~(\ref{eq:Schwarz}), 
for a period of $10~\rm s$ (approximately 25 large eddy turnover times
of the normal fluid).
We find that, after an initial transient, the vortex line density
saturates to a statistically 
steady state, as shown in Fig.~\ref{fig1}, 
independently of the initial condition (various vortex loop configurations 
were tried).
\begin{figure}
\begin{center}
\includegraphics[
  width=0.8\linewidth,
  keepaspectratio]{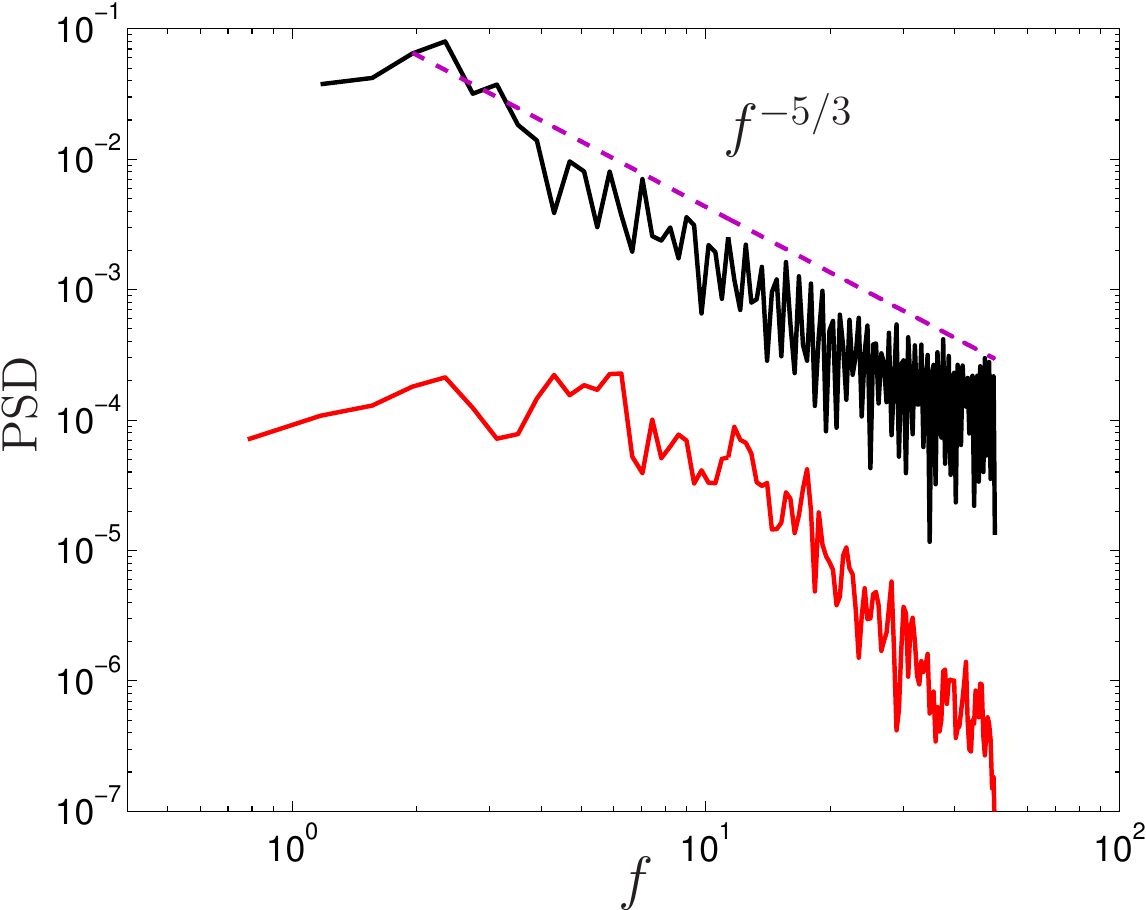}
\end{center}
\caption{
(Color online).
Power spectral density, PSD, (arbitrary units) of fluctuations of the total
vortex line density $L$ (black upper solid line)
and of the polarised vortex line density $L_{\parallel}$ (red lower solid line)
vs frequency $f$ ($\rm s^{-1}$). 
The dashed (magenta) line shows the $f^{-5/3}$ scaling. 
}
\label{fig2}
\vspace{-4mm}
\end{figure}
\begin{figure*}
\begin{center}
\includegraphics[
  width=0.7\linewidth]{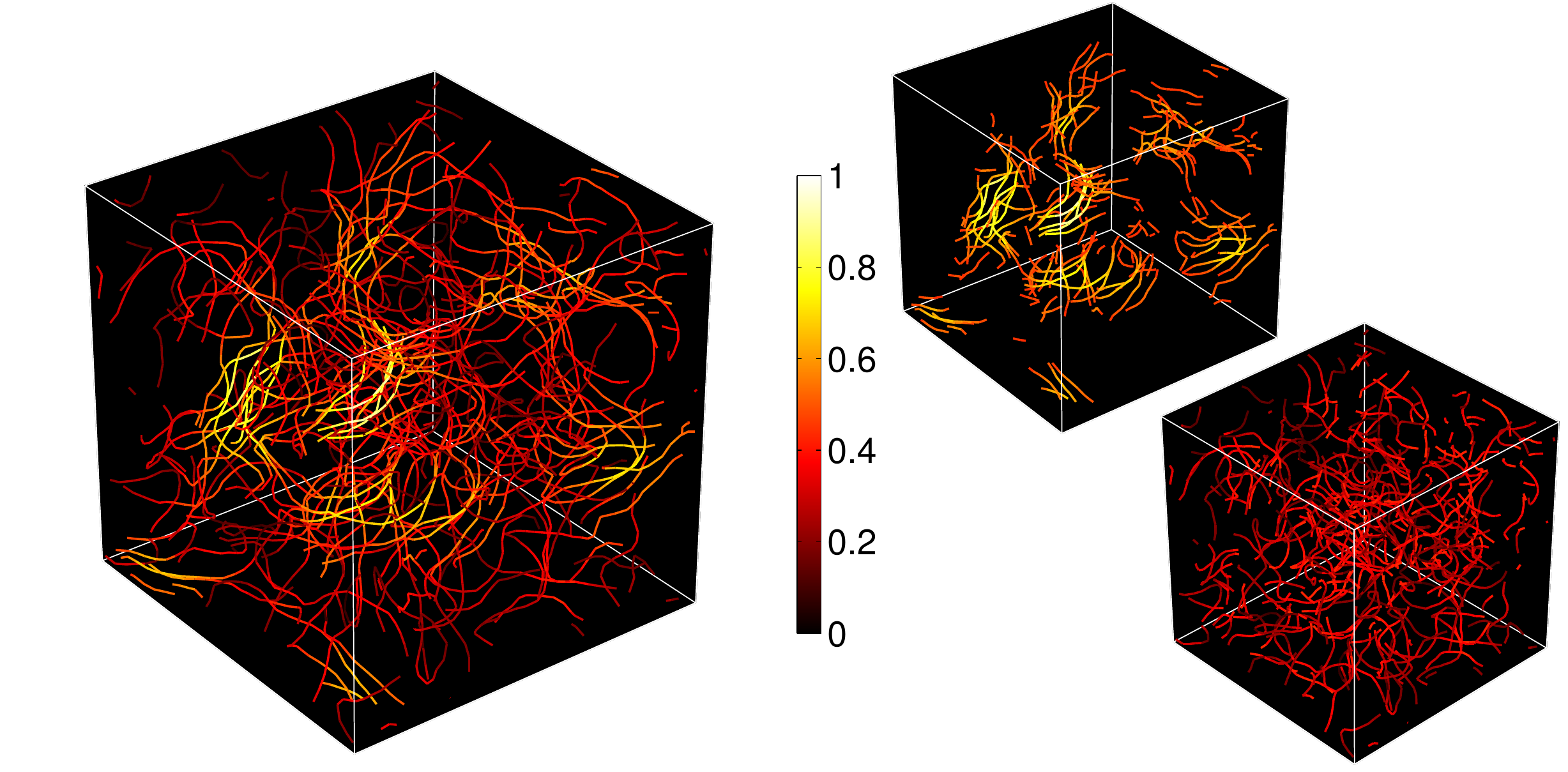}
\end{center}
\caption{(Color online) 
Left: Snapshot of the vortex tangle at $t=7\,$s. Vortex lines are
locally coloured according to the local magnitude of the smoothed vorticity 
field ${\bom}$. Right: The same snapshot but split into the locally 
polarized (top, ${\omega}(\bs_i)>1.4\omega_{\rm rms}$) and random components 
(bottom, ${\omega}(\bs_i)<1.4\omega_{\rm rms}$) respectively. 
The color scale is normalised by the maximum value 
$\omega_{\rm max}=38.4\,$s$^{-1}$; $\omega_{\rm rms}=15.7\,$s$^{-1}$}.
\label{fig3}
\vspace{-8mm}
\end{figure*}

\begin{figure}
\begin{center}
\includegraphics[%
  width=0.7\linewidth,
  keepaspectratio]{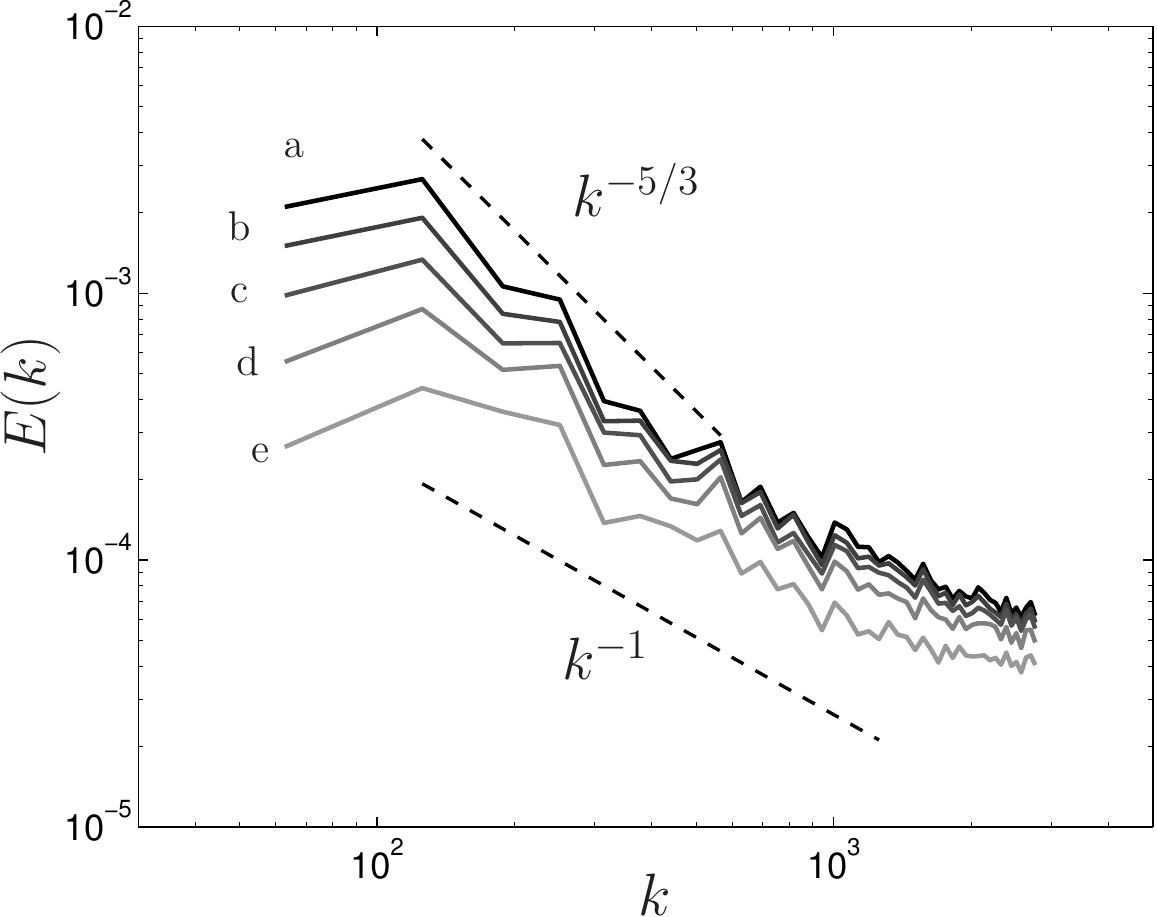}
\includegraphics[
width=0.7\linewidth,
  keepaspectratio]{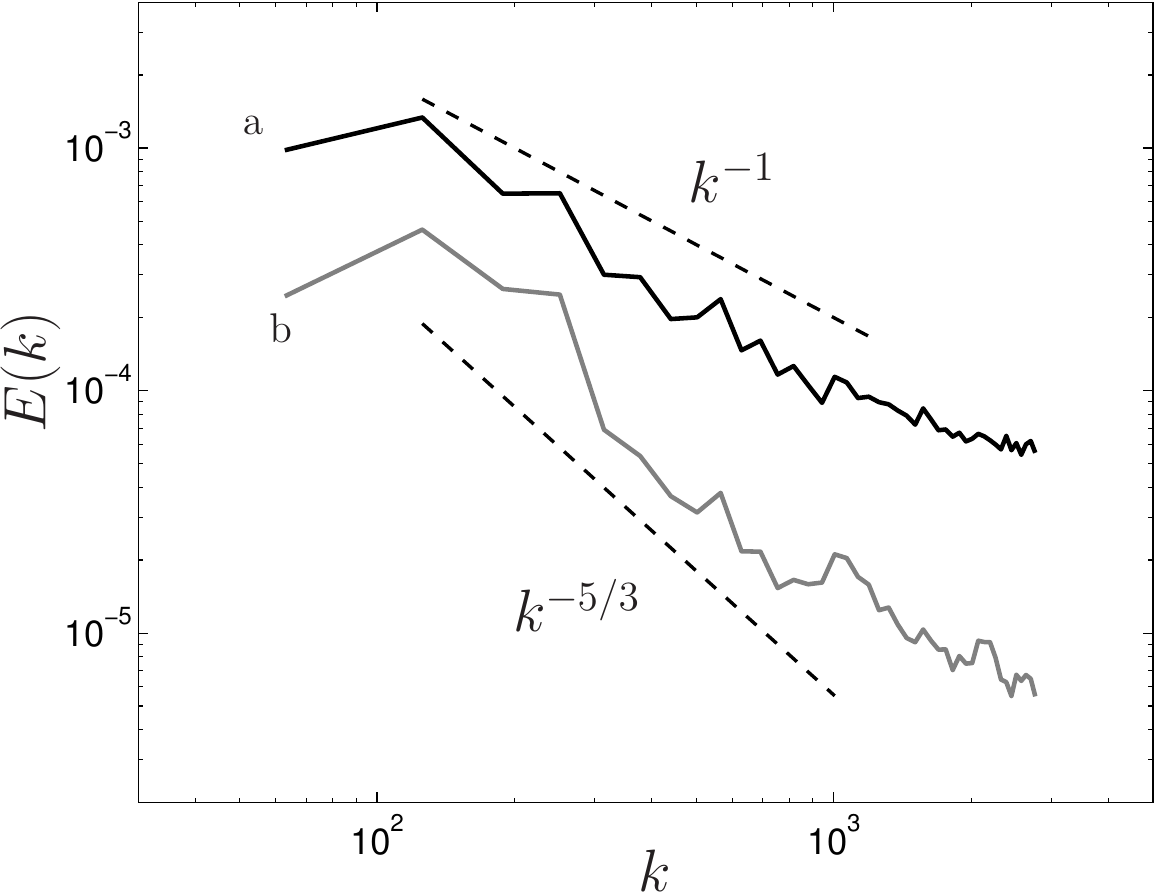}
\end{center}
\caption{
Energy spectra $E(k)$ (arbitrary units) vs wavenumber $k$ ($\rm cm^{-1}$)
corresponding to Fig.~\ref{fig3}. 
Top: The upper solid line (a) is the energy spectrum corresponds to the
flow induced by all vortex lines. 
Lower curves correspond 
to the flow induced by vortices with smoothed vorticity below 
the following thresholds: 
(b) $\omega<1.7\omega_{rms}$, 
(c) $1.4\omega_{rms}$, 
(d) $1.2\omega_{rms}$, and
(e) $\omega_{rms}$. 
Bottom: Energy spectra corresponding to
vortex lines with smoothed vorticity respectively below (a) and above (b) 
the threshold $1.4\omega_{rms}$. The dashed lines display the
$k^{-1}$ and the $k^{-5/3}$ scalings. 
}
\label{fig4}
\vspace{-4mm}
\end{figure}

We analyse the properties of the superfluid turbulence
in the statistically steady state ($t\gtrsim 1.2\,$s). 
Firstly, we compute the 
frequency spectrum of the fluctuations of the vortex line density about
its average value $<L>=1.15 \times 10^4 ~\rm cm^{-2}$. 
Fig.~\ref{fig2} shows that the spectrum scales as $f^{-5/3}$ 
for large $f$, as observed in experiments 
\cite{Roche2007,Bradley2008}
and numerical simulations \cite{Baggaley-fluctuations}.
Secondly, we compute the superfluid energy spectrum $E(k)$, defined by
\begin{equation}
\frac{1}{|V|}\int_V \frac{1}{2} \vert {\bf v}_s \vert^2 d{\bf x}=\int E(k) dk,
\label{eq:kspectrum}
\end{equation}

\noindent
where $V=D^3$ is volume.  The top curve (labelled a)
of Fig.~\ref{fig4} shows that the energy spectrum 
(computed on a $256^3$ Cartesian mesh) is consistent with
the cascading classical Kolmogorov scaling $E(k) \sim k^{-5/3}$ in the range
$k_D=2 \pi/D < k < k_{\ell}=2 \pi /{\ell}$, in agreement
with experiments \cite{Maurer,Salort} and numerical simulations
\cite{Araki,Baggaley-fluctuations,Nore,Kobayashi}; at larger wavenumbers
$k>k_{\ell}$ the spectrum exhibits the non-cascading scaling of
individual vortex lines. We conclude that the numerical model 
reproduces the two major spectral features (the energy spectrum and
the vortex density fluctuations spectrum) observed in superfluid turbulence.

To solve the puzzle described in the introduction, we examine
the homogeneity of the turbulence. A simple three-dimensional plot of the
vortex lines may give the wrong impression
that the vortex tangle is spatially uniform: the orientation
of the lines is not apparent and some lines
partially hide each other. A more careful analysis is required.
We define a smoothed vorticity field $\bom$ 
at the discretization points $\bs_j$ using a kernel with finite support, 
the $M_4$ kernel~\cite{Monaghan-Review} (effectively a cubic spline):

\begin{equation}
\bom(\bs_i)=\kappa \sum_{j=1}^N \bs'_j W(r_{ij},h) \Delta \xi_j,
\label{eq:smooth}
\end{equation}

\noindent
where $r_{ij}=|\bs_i-\bs_j|$, $\Delta \xi_j=|\bs_{j+1}-\bs_j|$,
$W(r,h)=g(r/h)/(\pi h^3),$ 
$h$ is a characteristic length scale, and

\begin{equation}
g(q) =
\begin{cases}
1 - \frac{3}{2}q^{2} + \frac{3}{4}q^{3} , & 0 \leq q < 1 ; \\
\frac{1}{4}\left( 2 - q\right)^{3}, & 1 \leq q < 2 ;\\
0, & q \geq 2.
\end{cases}
\label{eq:spline}
\end{equation}

\noindent
This approach is commonly employed in the smoothed particle hydrodynamics 
(SPH) literature~\cite{Monaghan-Review},
as we only need to take the contribution of discretization
points within a 
radius of $2h$ from each point $\bs_i$. 
We tested this smoothing algorithm against a Gaussian 
kernel $g(q)=e^{-q^2}$ \cite{Baggaley-structures} and found comparable 
results for single snapshots of the vortex tangle. The advantage of
the new algorithm is that $\bom$ is computed only at the discretization
points, so it can be evaluated during the time evolution. 

Setting $h=\ell$, the smoothed vorticity field $\bom$
allows us to identify the presence of any coherent vortex structures.
From the vorticity values $\bom_j$ at each discretization point $\bs_j$ 
we compute the rms vorticity
$
\omega_{\rm rms}=\sqrt{\frac{1}{N}\sum_{i=1}^N |{\bom}(\bs_i)|^2}.
$
and, following Ref.~\cite{Roche-Barenghi},
we decompose the vortex lines into locally `polarised' and
`unpolarised` fields, $L_{\parallel}$ and $L_{\times}$ respectively.
The polarised vortex line density
$L_{\parallel}$ consists of the discretization
points $\bs_j$ associated to intense metastable vortical regions where the
magnitude of the smoothed vorticity exceeds the rms vorticity by
a threshold value,
$\omega(\bs_i)>1.4\omega_{\rm rms}$ (we shall discuss this threshold
shortly). The remaining discretization
points form the `unpolarised' field $L_{\times}$. As with the total vortex line density, we find that $L_{\parallel}$ rapidly saturates to a statistically steady state, see Fig.~\ref{fig1}.

A snapshot of the vortex tangle in which the vortex lines
are coloured according to the local value of the smoothed vorticity 
is shown in Fig.~\ref{fig3} (left); vortex bundles are clearly visible.
The right-hand side of Fig.~\ref{fig3} shows respectively the polarised 
(top, $\omega(\bs_j)>1.4 \omega_{\rm rms}$) and  unpolarised
vortex lines (bottom, $\omega(\bs_j)<1.4\omega_{\rm rms}$): the former
consists of bundles and contributes to $L_{\parallel}$, the latter 
is spatially random and contributes to $L_{\times}$. 

We find that the polarised and unpolarised vortex lines have different 
properties.  The frequency spectrum of the fluctuations of $L_{\times}$
scales as $f^{-5/3}$, as for the total line density $L$ 
(top black curve of Fig.~\ref{fig2}), whereas
the spectrum of the fluctuations of 
$L_{\parallel}$ slowly increases at low frequency and drops rapidly at high
frequency (bottom red curve of Fig.~\ref{fig2}). Changing the threshold
value within the range 1.4 to 2.2$\omega_{\rm rms}$ does not change this
distinction.
The flattening and the slow increase with $f$ of the spectrum of the
fluctuations of the polarised vortex lines are consistent with observations in ordinary turbulence~\cite{Ishihara,Zhou}.

We now turn our attention to the role of the polarised lines on the energy
spectrum.  We have seen that the top curve (labelled a) 
of the top panel of Fig.~\ref{fig4} 
displays, in the range $k_D<k<k_{\ell}$, 
the Kolmogorov energy spectrum corresponding to all vortex lines.
The curves labelled b, c, d, and e 
show energy spectra arising only from discretization
points ${\bs_j}$ such that $\omega$ is
below a given (decreasing) rms threshold.
It is clear that, as we remove the contribution of the 
high-intensity bundles, the energy spectrum $E(k)$ in the range
$k_D<k<k_\ell$ becomes shallower, and changes from a $k^{-5/3}$ scaling
to a $k^{-1}$ scaling. The bottom panel of Fig.~\ref{fig4} compares the 
energy spectrum arising from the isotropic vorticity $L_{\times}$ 
(top curve, labelled a) and 
that arising from the  polarised vorticity $L_{\parallel}$
(bottom curve, labelled b). It is apparent that the vortex bundles
correspond to the classical $k^{-5/3}$ Kolmogorov scaling, and that
the random vorticity corresponds to the $k^{-1}$ spectrum.

The decomposition of $L$ into $L_{\parallel}$ and $L_{\times}$
is robust and holds during the time evolution, thus vindicating
Roche and Barenghi \cite{Roche-Barenghi} who suggested it when
discussing an experiment \cite{Roche2007}. Unlike that experiment, in
our calculation the random vortex lines contain most of the energy.
A possible explanation of this difference is that our model of 
turbulent normal fluid does not contain strong vortical structures, 
unlike real turbulence or DNS, thus our polarised bundles are an 
underestimate of reality 
(the presence of vortical structures in the normal fluid 
would certainly induce stronger vortex bundles in the superfluid, as shown in
numerical simulations \cite{Koplik-Morris}, hence
increase the energy contained in $L_{\parallel}$).  
It is also interesting to remark that the the smoothed vorticity
$\bom$  can be related to the coarse-grained vorticity 
of the HVBK equations \cite{HVBK}, and that
the decomposition of $L$ into $L_{\parallel}$ and $L_{\times}$,
has an analogy to that of Farge et al. \cite{Farge}
for classical turbulence.

In summary, we have shown compelling evidence that, at any instant,
the turbulent vortex tangle can be decomposed into a polarised and a random
component.
The polarised component, associated to $L_{\parallel}$, 
consists of metastable coherent vortex bundles, shown in the top right panel of Fig.~\ref{fig3}, and is responsible for the
observed Kolmogorov $k^{-5/3}$ energy spectrum.
The random component, shown in the bottom
right panel of Fig.~\ref{fig3}, is responsible for the observed
$f^{-5/3}$ frequency spectrum of the fluctuations of the vortex length.
The result confirms that the Kolmogorov spectrum arises from partial
polarization of the vortex lines, and solves an apparent puzzle 
between experiments. It also points to direction of further work:
determining the degree of polarization as a function of temperature
and normal fluid's Reynolds number (in Fig.~\ref{fig1} it is
about 20 percent with the parameters used),
and developing a two-scale approach to superfluid hydrodynamics
such as Lipniacki's \cite{Lipniacki} to account for such polarization.

This work was supported by the Leverhulme Trust, grant numbers F/00 125/AH 
and F/00 125/AD, and the ANR program STATOCEAN (ANR-09-SYSC-014). 
AWB acknowledges helpful discussions with Paul Clark.

\end{document}